\documentclass[12pt]{article}
\usepackage[utf8]{inputenc}
\usepackage{amsmath}
\usepackage{graphicx}
\usepackage{amssymb}
\numberwithin{equation}{section}
\usepackage{mathtools}
\usepackage{physics}

\usepackage[page, toc]{appendix}

\usepackage				[usenames,dvipsnames]	{color} 
\usepackage										{textcomp}
\usepackage										{leftidx}
\usepackage										{ulem}
\usepackage										{bbding}
\usepackage										{diagbox}
\usepackage										{multirow}
\usepackage										{longtable}
\usepackage										{listings}
\usepackage										{xcolor}
\usepackage										{pdflscape}
\usepackage										{bbm}
\usepackage										{amsbsy}
\usepackage										{pdfpages}
\usepackage										{subcaption}
\usepackage										{siunitx}
\usepackage										{float}
\usepackage										{array}
\usepackage	[notextcomp, notext, nomath, not1]	{stix}
\usepackage										{dsfont}
\usepackage										{bigints}
\usepackage										{amsthm}
		\newtheorem{theorem}{Theorem}[section]
		\newtheorem*{theorem*}{Theorem}
		
		\newtheorem*{conjecture*}{Conjecture}

	\theoremstyle{definition}

	\theoremstyle{remark}

	\theoremstyle{plain}
		\newtheorem{example}{Example}[subsection]
		
\numberwithin{table}{section}

{
}

\usepackage										{jheppub}
\definecolor{darkblue}{rgb}{0.0,0.0,0.4}
\definecolor{darkgreen}{rgb}{0.0,0.4,0.0}
\hypersetup{
	colorlinks,
	linkcolor=black,
	citecolor=darkgreen,
	urlcolor=darkblue
}

\def\black{\color{black}}

\def\white{\color{white}}

\def\CC{{\mathbb C}}

\def\NN{{\mathbb N}}

\def\PP{{\mathbb P}}
\def\QQ{{\mathbb Q}}

\def\Cc{{\mathcal C}}

\def\Mm{{\mathcal M}}

\def\Oo{{\mathcal O}}

\def\Ss{{\mathcal S}}
\def\Tt{{\mathcal T}}

\def\Zz{{\mathcal Z}}

\def\bb{{\mathfrak{b}}}

\def\zz{{\mathfrak{z}}}

\def\sS{{\mathfrak{S}}}

\def\eps{{\varepsilon}}

\def\Mgn{{\Mm_{g, n}}}
\def\Mgnbar{{\overline{\Mm}_{g, n}}}
\def\M[#1, #2]{{\Mm_{#1, #2}}}
\def\Mbar[#1, #2]{{\overline{\Mm}_{#1, #2}}}
\def\Mc[#1, #2, #3]{{\Mm_{#1, #2}^{#3}}}
\def\Mbarc[#1, #2, #3]{{\overline{\Mm}_{#1, #2}^{#3}}}

\def\@KdV{{|_{\text{\textnormal{KdV}}}}}

\def\Res[#1]{{\underset{#1}{\text{\textnormal{Res }}}}}

\def\W[#1, #2]{{W_{#2}^{(#1)}}}

\def\bbrack[#1]{{[\![#1]\!]}}
\def\angles[#1, #2]{{\langle #1 \rangle_{#2}}}
\def\sumkappa[#1]{{e^{\sum_k \tilde{t}_{a_{#1}, k} \kappa_k}}}

\def\spec[#1]{{\text{Spec}(#1)}}

\def\ubraced[#1, #2]{{\underset{#2}{\underbrace{#1}}}}

\def\out[#1]{{}}

\date{date}

\begin{document}

\baselineskip=15pt
\begin{titlepage} 
\begin{center}
\vspace{1cm}   
\vspace*{ 2.0cm}
{\Large {\bf Intersection theory of the complex quartic Kontsevich model}}\\[12pt]
\vspace{-0.1cm}
\bigskip
\bigskip 
{ {{Finn Bjarne Kohl}$^{\,\text{a, b}}$}, {{Raimar Wulkenhaar}$^{\,\text{b}}$}
\bigskip }\\[3pt]
\vspace{0.cm}
{
  ${}^{\text{a}}$ 
  {\it
  	Institute for Theoretical Physics,~Westfälische Wilhelms-Universität Münster, Wilhelm-Klemm-Straße 9, 48149 Münster, Germany\\
  }
  ${}^{\text{b}}$ 
  {\it
  	Mathematical Institute,~Westfälische Wilhelms-Universität Münster,\\ Einsteinstraße 62, 48149 Münster, Germany\\
  }
}
\vspace{2cm}
\end{center}

\begin{abstract}
\noindent We expand correlation functions of the Langmann-Szabo-Zarembo (LSZ) model in terms of intersection numbers on the moduli space of complex curves. This provides an explicit, physically motivated example for the expansion of correlation functions generated by Chekhov-Eynard-Orantin topological recursion. To this end, we unify notation as well as different conventions present in the literature and use a set of moduli of the spectral curve adapted to the physically motivated model. The presentation focuses on an illustrative, step-by-step comprehension of the work.
\end{abstract}

\end{titlepage}
\clearpage
\setcounter{footnote}{0}
\setcounter{tocdepth}{2}

\tableofcontents

\newpage 
\section{Introduction}
Topological recursion has been developed by L. Chekhov, B. Eynard, and N. Orantin formalizing the solution strategy of certain matrix models \cite{Chekhov_2006a, Chekhov_2006b, Chekhov_2006c, Eynard:2007kz}. This technique, which turned out to be a universal structure underlying to various problems, uses complex analytic tools to recursively compute essential objects called correlation functions from initial data provided by a spectral curve. The theory is embedded in a web of mathematical fields connected by deep theorems since the 90s \cite{Belliard21}.\\
Then M. Kontsevich \cite{Kontsevich:1992ti} proved E. Wittens conjecture \cite{Witten:1990hr} relating intersection numbers on the moduli space of complex curves to a special class of integrable hierarchies. In his seminal work he interpreted the moduli problem first in terms of enumerative geometry. Then used complex analytic tools, which can in modern language be formulated in terms of topological recursion, to apply work of Dijkgraaf, Verlinde and Verlinde \cite{DIJKGRAAF1991435} to prove integrability. \\
Beyond this prime example, there are various cases where the correlation functions computed from topological recursion have enumerative geometric interpretation. Furthermore, it is generally provided that theories obeying topological recursion correspond to a solution for a class of integrable hierarchies \cite{Eynard:2007kz}.\\
In this work the relation of topological recursion to the moduli space of complex curves is studied, based on work of B. Eynard \cite{Eynard1104, Eynard1110}. There he proves the expansion of correlators generated by topological recursion from regular spectral curves with finitely many ramification points in terms of intersection numbers on the moduli space of curves. We provide an illustrative step-by-step example expanding correlators of a physically motivated theory. By that means we intend not only to give more insight into the structures involved but also to unify notation as well as conventions.\\

Quantum field theories provide a proper framework to describe observations in particle physics since their introduction in the beginning of the past century. Despite its great success, its formulation is plagued with a number of issues and remains a topic of active research in physics and mathematics throughout the decades. Inspired by the ambition to find a quantum theory of gravity in the beginning of this century the study of quantum field theories on non-commutative space has received attention. Prominent instances are the Grosse-Wulkenhaar model a theory of hermitian fields with quartic interaction \cite{GW05, Grosse:2012uv} as well as the Langmann-Szabo-Zarembo (LSZ) model \cite{Langmann_2004}.\\
Recently significant progress has been made by the solution of the LSZ model with quartic interaction by Branahl and Hock in \cite{BH2205}. In their work the authors are able to interpret Dyson-Schwinger equations for correlators of the model as abstract loop equations. The analysis of their pole structure allowed Branahl and Hock to show that the correlators of the LSZ model follow Chekhov-Eynard-Orantin topological recursion \cite{Eynard:2007kz}.\footnote{A similar analysis is conjectured \cite{HW2103, BHW2208} to yield the solvability of the Grosse-Wulkenhaar model by Borot-Shadrin blobbed topological recursion \cite{Borot:2015hna}.} In \cite{BH2205} the authors provide the information about the spectral curve, which constitutes the central object of the analysis in this work. \\

In the following we are going to describe the necessary parts of the theory of topological recursion and of intersection numbers on the moduli space of complex curves (cf. Sections~\ref{sec:topological-recursion} and \ref{sec:intersection-numbers}). Thereafter, the expansion of correlation functions generated by topological recursion in terms of intersection numbers is recalled, following Eynard~\cite{Eynard1110} (cf. Section~\ref{sec:TR-IN}) and in terms of different expansion data (cf. Section~\ref{sec:XiYi}). These sections should provide a unified notation that can be used in the later analysis. In Section~\ref{sec:LSZ} we give a short reminder on the model that is investigated in this work and comment on its combinatorial limit. Details and results of the analysis can be found in Section~\ref{sec:results}, with the data listed in Appendix~\ref{app:complexmodel-expansiondata-xiyi}. 

\section{Topological recursion}\label{sec:topological-recursion}
In the beginning of the 2000s the study of matrix models culminated in the introduction of topological recursion in \cite{Eynard:2007kz} as a central framework that computes the correlators of the system recursively. The subsequent work generalized this framework immensely and yielded numerous applications throughout mathematics and physics \cite{Eynard:2016yaa, Belliard21}. Proving topological recursion for a specific problem/model is in general a hard task and has crucial implications. Importantly, it implies exact solvability and integrability in terms of some associated integrable hierarchy. Most often it is revealed that certain physically motivated models correspond to specific enumerative problems and vice versa. The task of showing Eynard-Orantin topological recursion for the LSZ model (see Section~\ref{sec:LSZ}) was completed in \cite{BH2205}. This extends the proof of exact solvability of the original work and, importantly, establishes similarities to other models such as the Grosse-Wulkenhaar model.

\begin{figure}[t]
	\centering
	\includegraphics[width=\linewidth]{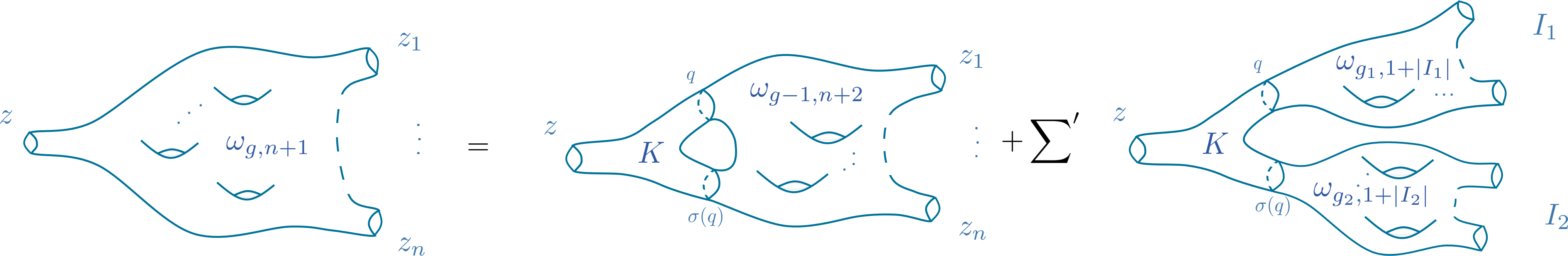}
	\caption{Illustration of equation~\eqref{equ:TR}. According to decompositions in terms of pairs of pants a Riemann surface of type $(g, n+1)$ can be found from glueing one of type $(g-1, n+2)$ or two of type $(g_1, 1+\abs{I_1})$ and $(g_2, 1+\abs{I_2})$. Note that topological recursion is \textit{not} associated to a decomposition in terms of pairs of pants but pairs of pants and cylinders, i.e. surfaces of topology $(0, 2)$ and $(0, 3)$. }
	\label{fig:trillustration}
\end{figure}

In the following, the original Eynard-Orantin topological recursion is reviewed. For details the reader is referred to \cite{Eynard:2007kz} or various later works reviewing and generalizing the framework. \\
Topological recursion computes an infinite sequence of admissible correlators $\{\omega_{g, n}\}_{g, n}$ indexed by $g\in \NN$, and $n\in \NN^\times$, which are meromorphic 1-forms in each of their $n$ variables and are defined on an $n$-fold product of a Riemann surface $\Sigma$ with genus $g$. The recursion is conducted in $(2(g-1)+n)$ and the initial data is provided by a spectral curve $\Ss$. It consists of a Riemann surface $\Sigma$ as well as two meromorphic functions $x$ and $y$, which are maps from $\Sigma$ to a base Riemann surface $\Sigma_0$ . \\
The first correlator is, then, given by 
\begin{align}
	\omega_{0, 1}(z_1) = y(z_1)\dd{x(z_1)}\, .
\end{align}
Note that only regular spectral curves are considered here. This requires the zeros of $\dd{x}$, the ramification points $\{a_i\}_{i=1, \dots \bb}$, to be simple as well as $y$ and $\dd{y}$ is non-vanishing there. Furthermore, the so-called Bergman kernel $B$, which is defined to be a normalized, symmetric second-kind bilinear meromorphic differential, having a double pole on the diagonal and no other pole, gives the correlator
\begin{align}
	\omega_{0, 2}(z_1, z_2) = B(z_1, z_2) \overset{z_1\to z_2}{\sim} \frac{\dd{z_1}\dd{z_2}}{(z_1-z_2)^2}+ \text{holomorphic}\, .
\end{align}
At this point all "unstable" correlators are provided by the spectral curve $\Ss=(\Sigma, x, y, B)$. The recursion (also illustrated in Figure~\ref{fig:trillustration})
\begin{align}\label{equ:TR}
	\omega_{g, n+1}(z, I) = \sum\nolimits_{i=1}^\bb \Res[q\to a_i] &\; K_i(z, q) \left[ \omega_{g-1, n+2}(q, \varsigma_i(q), I)\white \sum_m^m \right.\nonumber\\
	& \left. + \sum_{g_1+g_2=g}\sum_{I_1\sqcup I_2=I}^{\prime} \omega_{g_1, 1+\abs{I_1}}(q, I_1) \; \omega_{g_2, 1+\abs{I_2}}(\varsigma_i(q), I_2)\right]
\end{align}
computes all higher correlators, which are meromorphic with poles only at the ramification points with vanishing residue, where $\varsigma_i$ is the local involution of $x$ at the $i$'th marked point. 
These correlators obey so-called abstract loop equations \cite{Borot:2013lpa}, however, do not provide the most general solution to the latter \cite{Borot:2015hna}. In the recursion above the set $I$ is given by $I=\{z_1, \dots, z_n\}$ and in the primed sum terms containing $(0, 1)$-type correlators are excluded. The recursion kernel $K_i$ is composed of the initial data as
\begin{align}
	K_i(z, q) = \frac{\frac{1}{2}\int_{\varsigma_i(q)}^{q}B(z, \bullet)}{\omega_{0, 1}(q)-\omega_{0, 1}(\varsigma_i(q))}\, .
\end{align}

\newpage
\section{Intersection numbers on moduli spaces}\label{sec:intersection-numbers}

Initiated by the work of M. Kontsevich \cite{Kontsevich:1992ti}, B. Eynard was able to relate the correlation functions $\omega_{g, n}$ computed by topological recursion to intersection theory on the moduli space of complex curves \cite{Eynard1104, Eynard1110}.\\
In order to provide a brief reminder (following \cite{Eynard1110}), let $\Mgn$ be the moduli space of smooth irreducible complex curves of genus $g$ with $n$ marked points, which is a complex orbifold of dimension $\dim_\CC \Mgn = 3(g-1)+n$. As families of smooth curves may degenerate into non-smooth curves, $\Mgn$ is not compact. Its well known Deligne-Mumford compactification $\Mgnbar$, containing also stable nodal curves\footnote{The stability condition can be expressed as the finiteness of the automorphism group of the respective complex curve and is fulfilled if and only if for every irreducible component $\chi=2(1-g)-n<0$. In the vicinity of a nodal singularity the curve is isomorphic to $\CC\bbrack[x, y]/(xy)$.}, fixes this. \\
Intersection numbers on $\Mgnbar$ are defined as integrals of cohomology classes. In order to make this precise, let 
\begin{align}
	\pi:\; \Mbar[g, n+1]\to \Mbar[g, n]
\end{align}
be the forgetful morphism, under which the information about the $(n+1)$'st marked point is lost. Furthermore, denote by $s_i$ the $i$'th canonical section of $\pi$ and by $D_i$ the associated divisor, for $1\leq i \leq n$, and $\omega_\pi$ the associated relative dualizing sheaf. Then the classes of interest in this article include products of  $\psi$-classes
\begin{align}
	\psi_i= c_1(s_i^*(\omega_\pi))\, ,
\end{align}
which are differential forms of complex degree one. In the smooth case these simplify to the first Chern class of co-tangent bundle at the $i$'th marked point.\\
Further, the structure is enriched by including
\begin{align}
	\kappa_k= \pi_*c_1\left(\omega_\pi \otimes \Oo\left(\sum\nolimits_i D_i\right)\right)^{k+1}\, ,
\end{align}
which are forms\footnote{for more details see \cite{Mumford1983, Arbarello:1994sda, LX09}} of complex degree $k$. Note that 
\begin{align}
	(\pi_{n+1})_* \left( \psi_1^{a_1} \cdots \psi_{n}^{a_{n}} \psi_{n+1}^{a_{n+1}+1} \right) =&\; \psi_1^{a_1}\cdots\psi_{n}^{a_{n}} \kappa_{a_{n+1}}\, ,\\
	(\pi_{n+1}\pi_{n+2})_* \left( \psi_1^{a_1}\cdots\psi_n^{a_n} \psi_{n+1}^{a_{n+1}+1} \psi_{n+2}^{a_{n+2}+1} \right) =&\; \psi_1^{a_1}\cdots\psi_n^{a_n} \left(\kappa_{a_{n+1}}\kappa_{a_{n+2}}+ \kappa_{a_{n+1}+a_{n+2}}\right)\, , 
\end{align}
and in general 
\begin{align}
	(\pi_{n+1}\cdots\pi_{n+k})_* \left(\psi_1^{a_1}\cdots\psi_n^{a_n} \psi_{n+1}^{a_{n+1}+1} \cdots\psi_{n+k}^{a_{n+k}+1}\right) =&\; \psi_1^{a_1}\cdots\psi_n^{a_n} \sum_{\tau \in \mathfrak{S}_{k}}\kappa_\tau\, ,
\end{align}
which might provide insight into the significance of $\kappa$-classes. In the above $\sS_{k}$ is the symmetric group of order $k$, and $\kappa_\tau = \kappa_\abs{\tau_1} \cdots \kappa_\abs{\tau_m}$ for the decomposition of $\tau$ into cycles~$\tau_i$, for $i=1, \dots, m$.\\
Intersection numbers on $\Mgnbar$ are then defined as integrals of products of powers of $\psi$- and $\kappa$-classes
\begin{align}
	\angles[(\dots), {g, n}] = \int_{\Mgnbar} (\dots)\, ,
\end{align}
if and only if the degree of the compound class matches the dimension of the moduli space (and zero otherwise).\\
In order to conveniently express the correlation functions associated to spectral curves with several branchpoints in terms of intersection numbers, following \cite{Eynard1110} it is useful to enrich the moduli space of complex curves by the information provided by some coloring map $\sigma$. Therefore, define
\begin{align}
	\Mbarc[g, n, \bb] = \left\{(\Cc; p_1, \dots, p_n; \sigma)\right\}\, ,
\end{align}
where $(\Cc; p_1, \dots, p_n)\in \Mgnbar$ and the coloring map 
\begin{align}
	\sigma:\; \Cc\backslash \{\text{nodes}\} \to \{1, \dots, \bb\}
\end{align}
is a continuous map for $\bb\in \NN^\times$. Note that continuity of $\sigma$ implies that it must be constant on each component of the curve $\Cc$.\\
In the two examples below $\Mbarc[g, n, \bb]$ is decomposed into factors of $\Mgnbar$, providing some idea of the structure of the enriched moduli space of colored complex curves. This should stress the fact that the colored moduli space is only an auxiliary construction abbreviating notation and structuring the calculations. All computations are equivalently done on (factors of the) ordinary $\Mgnbar$.
\begin{example}\label{ex:mbarc042}
	Consider $\Mbarc[0, 4, 2]$. Points in this space are colored curves of genus zero with four marked points as well as in its boundary nodal curves of topological type $(0, 3)\times (0, 3)$. As curves in the bulk only have one component (on which the color is supposed to be constant) and $\bb=2$, one finds two copies of $\M[0, 4]$ in $\Mbarc[0, 4, 2]$. Considering all possible colorings of boundary curves, one finds
	\begin{align}
		\Mbarc[0, 4, 2] =&\,  \M[0, 4] \cup \M[0, 4] \cup (\M[0, 3]\times\M[0, 3])^{\cup 12}\nonumber\\
		=&\, \Mbar[0, 4] \sqcup \Mbar[0, 4] \sqcup (\M[0, 3]\times\M[0, 3])^{\sqcup 6}\, .
	\end{align}
	Note that curves in $\Mbarc[0, 4, 2]$ of type $(0, 3)\times (0, 3)$, which are unichrome, are in the above decomposition in terms of \textit{compact} moduli spaces found in $\Mbar[0, 4]$. This gives rise to the different amount of factors of $(\M[0, 3]\times\M[0, 3])$ in the above (cf. Figure~\ref{fig:mbarc04-dec-coloring}).
	\begin{figure}[t]
		\centering
		\includegraphics[width=0.95\linewidth]{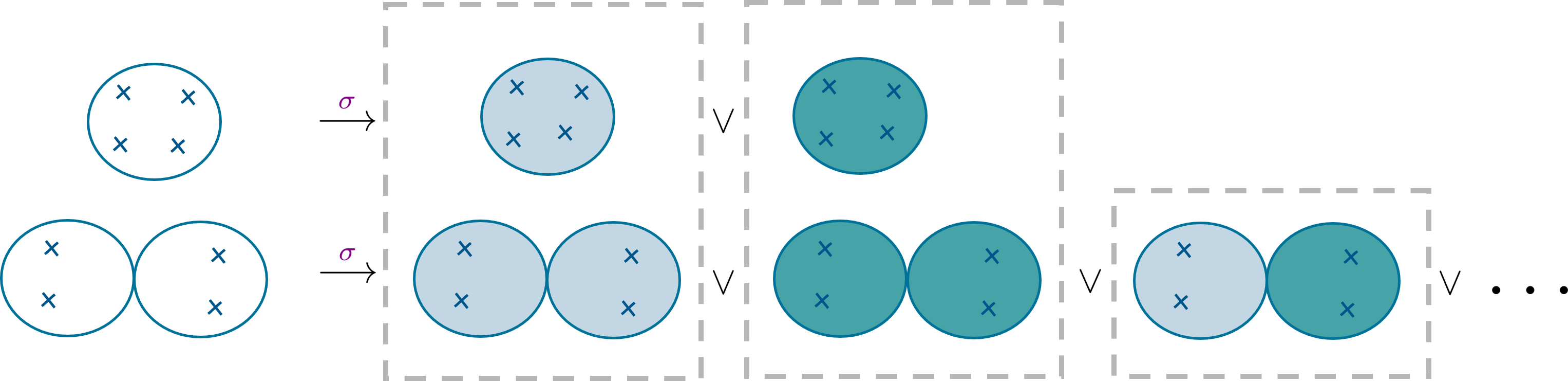}
		\caption{This depicts different colorings of curves of topological type $(0, 4)$ and $(0, 3)\times (0, 3)$, elements of $\Mbarc[0, 4, 2]$. Curves in the same gray box are in the same compact component of $\Mbarc[0, 4, 2]$. Note that the labelling of the marked points is omitted here to avoid cluttering.}
		\label{fig:mbarc04-dec-coloring}
	\end{figure}
\end{example}
\begin{example}\label{ex:mbarc112}
	As a second example take the decomposition of $\Mbarc[1, 1, 2]$. Curves of type~$(1, 1)$ in $\Mbarc[1, 1, 2]$ may degenerate into curves of type $(0, 3)$. As all curves have only one component, there are no polychrome curves. Thus, one finds
	\begin{align}
		\Mbarc[1, 1, 2]=&\; \M[1, 1] \cup \M[1, 1] \cup \M[0, 3] \cup \M[0, 3] \nonumber\\
		=&\; \Mbar[1, 1] \sqcup \Mbar[1, 1]\, .
	\end{align}
	An illustration can be found in Figure~\ref{fig:mbarc11-dec-coloring}.
	\begin{figure}[b]
		\centering
		\includegraphics[width=0.7\linewidth]{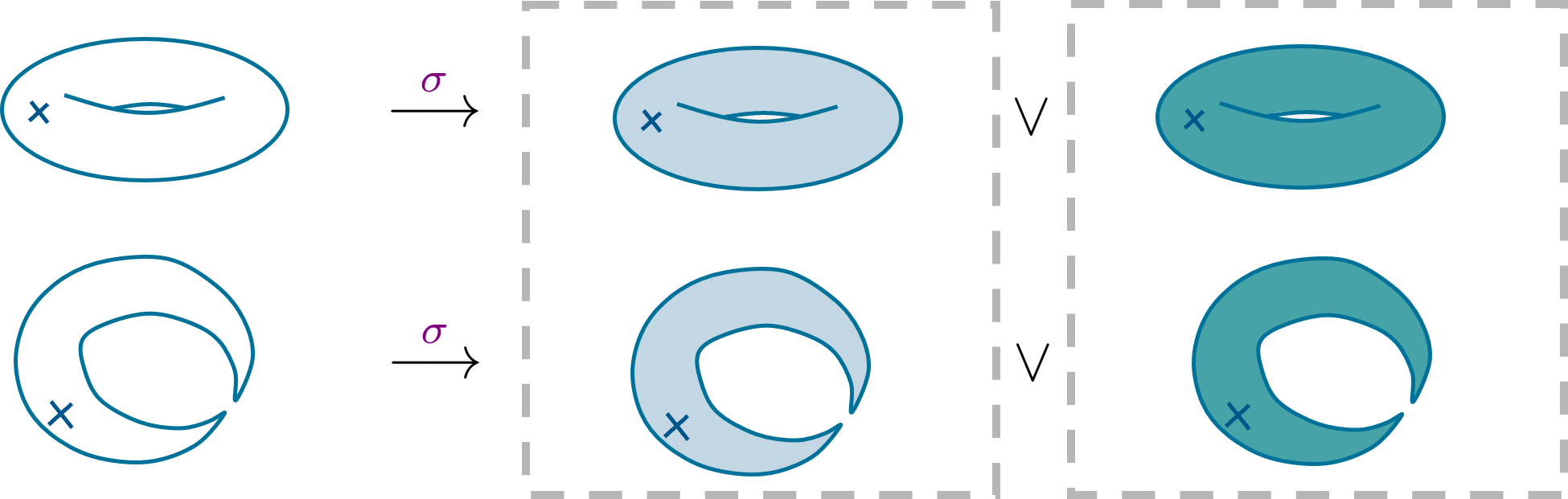}
		\caption{This depicts different colorings of curves of topological type $(1, 1)$ and $(0, 3)$, elements of $\Mbarc[1, 1, 2]$. Curves in the same gray box are in the same compact component of $\Mbarc[1, 1, 2]$. Note that the labelling of the marked points is omitted here to avoid cluttering.}
		\label{fig:mbarc11-dec-coloring}
	\end{figure}
\end{example}

\newpage
\section{Moduli space of curves and topological recursion}
Building on the work of M. Kontsevich \cite{Kontsevich:1992ti}, relating symplectic invariants associated to the Airy spectral curve to intersection numbers of $\psi$-classes, B. Eynard is able to show the expansion of the correlation functions generated by topological recursion in terms of $\psi$- and $\kappa$-classes. We cite his theorem here as Theorem~\ref{thm:main-thm}, presented in Subsection~\ref{sec:TR-IN}. His proof for spectral curves with one simple branchpoint relies on deformation arguments (cf. \cite{Eynard1104}). It is extended to spectral curves with finitely many simple branchpoints by glueing-arguments (cf. \cite{Eynard1110}).
\subsection{Topological recursion in terms of intersection numbers on $\Mgnbar$}\label{sec:TR-IN}
\begin{theorem}[Eynard '11]\label{thm:main-thm}
	Let $\Ss = (\Sigma; x, y; B)$ be a spectral curve with ramification points $\{a_1, \dots, a_\bb\}$ and consider local spectral curves $\Ss_i = (\Sigma_i; x, y; B_{\text{loc.}})$ where $\Sigma_i\subset \Sigma$ is a neighborhood of the ramification point $a_i$, $x$ and $y$ are the restrictions of $x$ and $y$ to $\Sigma_i$, and $B_{\text{loc.}}$ an arbitrary Bergman kernel on $\Sigma_i\times \Sigma_i$.\\
	Then, 
	\begin{align}\label{equ:main-thm}
		\omega_{g, n} (\Ss; z_1, \dots, z_n) = 2^{\dim_\CC \Mgn} \bigintsss_{\Mbarc[g, n, \bb]} \prod_{m=1}^{k} \hat{\Lambda}_{a_m} \prod_{(q_i, q_j)\in \{\text{nodes}\}} c(q_i, q_j) \prod_{i=1}^{n} \hat{B}_{a_{\sigma(p_i)}} (z_i; 1/\psi_i)\, .
	\end{align}
	While the left-hand-side is described in Section~\ref{sec:topological-recursion}, the right-hand-side is expressed in terms of intersection numbers and local expansion data of the spectral curve. The different contributions are introduced in the following sections.
\end{theorem}
Note that this theorem allows for a slightly more general setup in terms of the spectral curve. The spectral curve of the LSZ model is given in a global form and thus $B_{\text{loc.}}=B$. Furthermore, the combinatorial limit of the LSZ model considered here enforces $\bb=2$. The restriction to $\bb=2$ on the one hand simplifies the analysis immensely, but allows on the other hand still to observe phenomena that would not appear in the trivial case~$\bb=1$.
\subsubsection{Local expansion data of spectral curves}	
In the same manner as the correlation functions are computed in topological recursion through local residuum calculations, the data required to express these in terms of intersection numbers is extracted from local expansions of the spectral curve at the ramification points.\\
To that end, in the vicinity of a simple ramification point $a\in\{a_i\}_{i=1, \dots, \bb}$ define the local variable $\zeta_a$ in terms of the function $x$ by
\begin{align}
	\zeta_a(z)=\sqrt{x(z)-x(a)}
\end{align}
associated to the two sheets of the complex curve ramifying at $a$. In the following, the data, $y\dd{x}$ and $B$, defined on the spectral curve are expanded in $\zeta_a$ obtaining expansion data $t_{a, k}$ and $B_{a, k; a', k'}$. These encode as moduli the spectral curve $(\Sigma, y, x, B)$, a B-model geometry\footnote{The nomenclature of the "models" has been adopted in various works and originates in the string theory literature. In the context of topological recursion there is the notion of the "four sides" of a problem. A nice presentation of this can be found in \cite{Belliard21}. From this point of view what we call A-model here would correspond to the C-side and the B-model would be the B-side.}. B-model moduli are related through Laplace transformation to the moduli of the A-model, $\hat{t}_{a, k}$ and $\hat{B}_{a, k; a', k'}$. The transformation is given by a Laplace transformation in terms of $x$ and the integration path is the steepest descending contour $\gamma_a$ such that $x(\gamma_a) = \left[x(a), \infty \right[$.
\paragraph{Times $t_{a, k}$ } The \textit{times} $t_{a, k}$, defined as the expansion of $y(z)$ in terms of $\zeta_a$ as
\begin{align}\label{equ:times}
	y(z) \overset{z\to a}{=} \sum_{k\geq 0} t_{a, k+2} \, \zeta_a(z)^{k}\, ,
\end{align}
define the \textit{dual times} $\hat{t}_{a, k}$ through the Laplace transformation by 
\begin{align}\label{equ:t-tildet}
	\exp(-\sum_{k\geq0} \hat{t}_{a, k}u^{-k}) = 2 \sqrt{\frac{u}{\pi}} \int_{z\in\gamma_a} e^{-u \, \zeta_a(z)^2} \dd{y(z)}\, .
\end{align}
By inserting the expansion of $y$ and performing Gaussian integrals, one can establish the relation
\begin{align}\label{equ:t-ttilde}
	\sum_{k\geq 0} \hat{t}_{a, k}u^{-k} = -\log(2\sum_{k\geq 0} t_{a, 2k+3} \frac{(2k+1)!!}{2^k} u^{-k})\, ,
\end{align}
known as the Schur transform.
\paragraph{Coefficients $B_{a, k; a', k'}$ } Analogously, the Bergman-kernel is expanded in terms of the local variables
\begin{align}
	B(z, z') \overset{\substack{z\to a\\z'\to a'}}{=} \left(\frac{\delta_{a, a'}}{(\zeta_a(z)-\zeta_{a'}(z'))^2} + \sum_{k, k'\geq 0} B_{a, k; a', k'}\, \zeta_a(z)^k \zeta_{a'}(z')^{k'}\right) \dd{\zeta_a(z)} \dd{\zeta_{a'}(z')}\, .
\end{align}
The expansion coefficients $B_{a, k; a', k'}$ determine their duals defined by
\begin{align}\label{equ:hatB}
	\sum_{k, k'\geq 0} \hat{B}_{a, k; a', k'} \, u^{-k}v^{-k} =&\; \delta_{a, a'} \frac{uv}{u+v} + \frac{\sqrt{uv}}{2\pi} \int_{z\in\gamma_a}e^{-u \zeta_a(z)^2} \int_{z'\in\gamma_{a'}}e^{-v \zeta_a(z')^2} B(z, z')\nonumber\\
	=&\;\hat{B}_{a, a'}(u, v)\, .
\end{align} 
Expansion and evaluation of integrals yields
\begin{align}\label{equ:B-hatB}
	\sum_{k, k'\geq 0} \hat{B}_{a, k; a', k'} \, u^{-k}v^{-k} = \frac{1}{2}\sum_{k, k'\geq 0}B_{a, 2k; a', 2k'}\, \frac{(2k-1)!!}{2^k} \frac{(2k'-1)!!}{2^{k'}}\, u^{-k}v^{-k}\, .
\end{align}

Remark that from the relations~\eqref{equ:t-tildet} and \eqref{equ:B-hatB} it is explicit that the A-model moduli are determined by only half the B-model moduli. \\
The expansion data for the spectral curve associated to the LSZ model in the combinatorial limit with quartic interaction is provided in Appendix~\ref{app:complexmodel-expansiondata-xiyi}.
\subsubsection{Integration class}
\begin{figure}[t]
	\centering
	\includegraphics[width=0.5\linewidth]{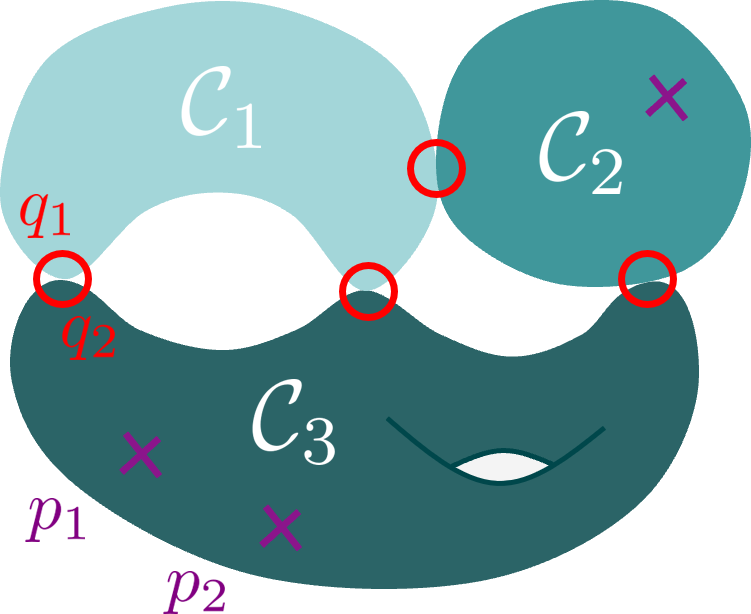}
	\caption{Sketch of an element of $\Mgnbar$, for $g=2$ and $n=3$. The different components are colored in different shades of green, the marked points $p_i$ are illustrated in violet and the nodes $(q_i, q_j)$ indicated in red.}
	\label{fig:integration-class-contributions}
\end{figure}

The local expansion data of the spectral curve described in the previous section provides all the information needed to define the integrand on the right-hand-side of equation~\eqref{equ:main-thm} in Theorem~\ref{thm:main-thm}.\\
The three factors of the integrand are due to contributions of elements in $\Mbarc[g, n, \bb]$ -- the bulk of the different components of a curve, its nodal points as well as the marked points (see Figure~\ref{fig:integration-class-contributions}). The product of these factors is integrated over the moduli space of colored complex curves. Comparing to the structure of $\Mbarc[g, n, \bb]$ explained in Section~\ref{sec:intersection-numbers} this amounts to an integration over the ordinary moduli space accompanied with a sum over colors.
\paragraph{Bulk contribution} Each component $\Cc_m$ of a colored curve $\Cc=(\cup_{m=1}^k\Cc_m) \in \Mbarc[g, n, \bb]$ contributes a factor
\begin{align}\label{equ:lambda-hat}
	\hat{\Lambda}_a= \exp(\sum_k\hat{t}_{a, k}\kappa_k + \frac{1}{2}\sum_{\delta\in\partial\Mgn}\sum_{k, k'} \hat{B_{\text{loc.}}}_{a, k; a, k'}{\ell_\delta}_* (\psi^k \psi'^{k'}))\, ,
\end{align}
where the sum $\sum_{\delta\in\partial\Mgn}$ runs over the co-dimension one boundary divisors $\delta$ of $\Mgn$. Note that at a boundary divisor the smooth curve degenerates yielding a node. Thus, a boundary divisor $\delta$ is isomorphic to either $\Mbar[g_1, n_1]\times\Mbar[g_2, n_2]$, with $g_1+g_2=g$ and $n_1+n_2=n+2$, or $\Mbar[g-1, n+2]$. For a fixed $\delta$ the map $\ell_\delta$ denotes the embedding 
\begin{align}
	\ell_\delta:\; \delta \hookrightarrow \Mgnbar\, .
\end{align}
By that means, the $\psi$-classes in equation~\eqref{equ:lambda-hat} are defined on either $\Mbar[g_1, n_1]\times\Mbar[g_2, n_2]$ or $\Mbar[g-1, n+2]$, where they are evaluated at the marked points, which make up the node on the respective component in $\partial\Mgnbar$. To be precise, for some class $\eta\in H^\bullet(\Mgnbar)$ one evaluates via the projection formula
\begin{align}\label{equ:projection-formula}
	\angles[\eta\; \ell_\delta{}_*(\psi^k\psi'^{k'}), {\Mgnbar}] = \angles[\ell_\delta{}^*(\eta) \; \psi^k\psi'^{k'}, \delta]\, .
\end{align}
If, further, $\delta\simeq \Mbar[g_1, n_1]\times\Mbar[g_2, n_2]$, one can decompose $\ell_\delta{}^*(\eta)=\sum_{i+j=\dim\delta-(k+k')} \rho_{1, i}\wedge\rho_{2, j}$ into components in $H^i(\Mbar[g_1, n_1], \QQ)\otimes H^j(\Mbar[g_2, n_2], \QQ)$, with $i+j=\dim\delta -(k+k')$ via the theorem by Künneth and find
\begin{align}
	\eqref{equ:projection-formula} \; = \sum_{i+j=\dim\delta-(k+k')} \angles[\rho_{1, i}\; \psi^k, {\Mbar[g_1, n_1]}]\angles[\rho_{2, j}\; \psi'^{k'}, {\Mbar[g_2, n_2]}]\, .
\end{align}
Note that due to the condition $\text{deg}(\text{integration class})=\dim(\text{moduli space})$ some terms in this sum will vanish.\\
It may also be remarked that $\hat{\Lambda}_a$ as well as $\hat{B_{\text{loc.}}}_{a, k; a, k'}$  is indexed by only one color. This is due to the fact that one component of a complex curve can be colored in just one single color (cf. Section~\ref{sec:intersection-numbers}).

\paragraph{Node contribution} 
The moduli space of colored complex curves contains components of polychrome nodal curves. These nodes contribute in the integral in equation~\eqref{equ:main-thm} in Theorem~\ref{thm:main-thm} through
\begin{align}\label{equ:c-node}
	c(q_i, q_j) = \sum_{d, d'\geq 0} \left(\hat{B}_{a_{\sigma(q_i)}, d; a_{\sigma(q_j)}, d'} - \delta_{a_{\sigma(q_i)}, a_{\sigma(q_j)}} \hat{B_{\text{loc.}}}_{a_{\sigma(q_i)}, d; a_{\sigma(q_j)}, d'} \right)\; \ell_* \left[\psi(q_i)^d\psi(q_j)^{d'}\right] \, .
\end{align}
The map $\ell$ here includes the boundary component that the considered nodal curve corresponds to into $\Mgnbar$. Its pushforward embeds the $\psi$-classes, which are evaluated on smooth marked points on a moduli space isomorphic to the boundary component, into $H^\bullet(\Mgnbar, \QQ)$.\\
Note that if $B=B_{\text{loc.}}$, only not self-connecting nodes give a non-zero result, that is those with $\sigma(q_i) \neq \sigma(q_j)$. This is also observed when decomposing $\Mbarc[g, n, \bb]$ into components of colored factors of $\Mgnbar$. Only \textit{polychrome} multi-component curves are due to novel components in $\Mbarc[g, n, \bb]$, while unichrome curves are parametrized by one of the $\bb$ factors of $\Mgnbar$ in terms of the disjoint decomposition of $\Mbarc[g, n, \bb]$ (cf. Example~\ref{ex:mbarc042} and \ref{ex:mbarc112}).

\paragraph{Contribution of marked points} In terms of graphs marked points correspond to half edges. This is reflected in the way marked points contribute in Theorem~\ref{thm:main-thm}. Each marked point contributes by a single sided Laplace transformation
\begin{align}\label{equ:b-marked-pts}
	\hat{B}_a(z; u) =&\; - \sqrt{\frac{u}{\pi}} \int_{z\in\gamma_a} e^{-u \; \zeta_a(z)^2} B(z, z')\nonumber\\
	=&\; \sum_{d\geq0} u ^{-d}\; \dd{\xi_{a, d}(z)}\, .
\end{align}
The expansion coefficients $\dd{\xi_{a, d}(z)}$ are one-forms and can be expanded as
\begin{align}
	\dd{\xi_{a, d}}(z) \overset{z\to a'}{=} \delta_{a, a'} \; \frac{(2d+1)!!}{2^d} \frac{\dd{\zeta_{a}(z)}}{\zeta_a(z)^{2d+2}}- \frac{(2d-1)!!}{2^d}\sum_{k\geq 0} B_{a, 2d; a', k}\;\zeta_{a'}(z)^k\dd{\zeta_{a'}(z)}\, .
\end{align}
Before comparing the left-hand-side and right-hand-side of equation~\eqref{equ:main-thm} in Theorem~\ref{thm:main-thm}, one needs to express both in terms of the same variables. The approach, followed here, is to evaluate the right-hand-side explicitly in terms of $\zeta$ and express the result then in terms of $z$. Therefore it is useful to express the basis of differential forms in terms of $\dd{z}$. This is conveniently done using 
\begin{align}
	\dd{\xi_{a, d}}(z) = -\frac{(2d-1)!!}{2^d} \; \Res[z'\to a] B(z', z)\; \zeta_a(z')^{-2d-1}\, .
\end{align}

In order to prepare the analysis described in Section~\ref{sec:results} two examples are provided. These can already be found in the original work \cite{Eynard1110}.
\begin{example}\label{eg:w04-1}
	Consider $\omega_{0, 4}$ for a spectral curve with $\bb=2$ and $B=B_{\text{loc.}}$. As described in example~\ref{ex:mbarc042}, the moduli space $\Mbarc[0, 4, 2]$ decomposes into two factors of $\Mbar[0, 4]$ and six factors of $\Mbar[0, 3]\times \Mbar[0, 3]$. Thus Theorem~\ref{thm:main-thm} expands to
	\begin{align}
		2^{-1}\omega_{0, 4}(z_1, \dots, z_4)&= \bigintsss_{\Mbar[0, 4]} \hat{\Lambda}_{a_1} \; \prod\nolimits_{i=1}^{4} \hat{B}_{a_1} (z_i; 1/\psi_i)+ \bigintsss_{\Mbar[0, 4]} \hat{\Lambda}_{a_2} \; \prod\nolimits_{i=1}^{4} \hat{B}_{a_2} (z_i; 1/\psi_i)\nonumber\\
		&\quad + \left(\bigintsss_{\Mbar[0, 3]\times\Mbar[0, 3]} \hat{\Lambda}_{a_1} \;\hat{\Lambda}_{a_2}\;c(q_1, q_2)\; \prod_{i=1}^{2} \hat{B}_{a_1} (z_i; 1/\psi_i) \prod_{j=3}^{4}\hat{B}_{a_2} (z_j; 1/\psi_j)\right. \nonumber\\
		&\qquad\qquad\qquad\qquad\qquad\qquad\qquad\left.\white \bigintsss \black  +\text{symm}(z_1, \dots, z_4)\right)\, .
	\end{align}
	Noting that $\dim\Mbar[0, 3]=0$ one expands the integrand in the second line according to definitions~\eqref{equ:lambda-hat}, \eqref{equ:c-node}, and \eqref{equ:b-marked-pts} keeping only those parts of degree zero. This yields for the integral over $\Mbar[0, 3]\times\Mbar[0, 3]$
	\begin{align}
		&\bigintsss_{\Mbar[0, 3]\times\Mbar[0, 3]} e^{\hat{t}_{a_1,0}\kappa_0}e^{\hat{t}_{a_2,0}\kappa_0} \; \hat{B}_{a_1, 0, a_2, 0}\, \psi(q_1)^0 \psi(q_2)^0\; \prod\nolimits_{j=1}^2 \dd{\xi_{a_1, 0}(z_j)}\psi_j^0 \prod\nolimits_{j'=3}^4 \dd{\xi_{a_2, 0}(z_{j'})}\psi_{j'}^0 \nonumber\\
		&= e^{(\hat{t}_{a_1,0}+\hat{t}_{a_2,0})\angles[\kappa_0, {0, 3}]}  \; \hat{B}_{a_1, 0, a_2, 0}\;  \angles[1, {0, 3}]\angles[1, {0, 3}]\; \dd{\xi_{a_1, 0}(z_1)} \dd{\xi_{a_1, 0}(z_2)} \dd{\xi_{a_2, 0}(z_3)} \dd{\xi_{a_2, 0}(z_4)}\, .
	\end{align}
	As $\dim\Mbar[0, 4]=1$, the integrand of the integral over $\Mbar[0, 4]$ is expanded to degree one. Note that with every order of $\hat{B_{\text{loc.}}}$ in the expansion of the bulk contribution $\hat{\Lambda}$ the dimension of the moduli space that is actually integrated over is reduced by one (cf. equation~\eqref{equ:projection-formula}). This yields
	\begin{align}
		& \bigintsss_{\Mbar[0, 4]} \left(e^{\hat{t}_{i, 0} \kappa_0} (1+\hat{t}_{i, 1}\kappa_1)\right) \left(1+\frac{1}{2}\sum\nolimits_{\delta\in\partial\Mbar[0, 4]}\hat{B}_{a_i, 0, a_i, 0}\psi^0\psi'^0\right) \prod\nolimits_{j=1}^{4} \left(\dd{\xi_{a_i, 0}(z_j)} \psi_j^0+\dd{\xi_{a_i, 1}(z_j)} \psi_j^1\right)\nonumber\\
		& =  \left( e^{\hat{t}_{a_i, 0} \angles[\kappa_0, {0, 4}]} \hat{t}_{a_i, 1} \angles[\kappa_1, {0, 4}] + e^{2\hat{t}_{a_i, 0} \angles[\kappa_0, {0, 3}]} \frac{6}{2}\hat{B}_{a_i, 0, a_i, 0}\; \angles[1, {0, 3}]\angles[1, {0, 3}]\right) \prod\nolimits_{j=1}^{4}\dd{\xi_{a_i, 0}(z_j)}\nonumber\\
		& \qquad + \left( e^{\hat{t}_{a_i, 0} \angles[\kappa_0, {0, 4}]} \; \angles[\psi_1, {0, 4}]\; \dd{\xi_{a_i, 1}(z_1)} \dd{\xi_{a_i, 0}(z_2)} \dd{\xi_{a_i, 0}(z_3)} \dd{\xi_{a_i, 0}(z_4)} + \text{symm}(z_1, \dots, z_4) \right)\, .
	\end{align}
	In conclusion
	\begin{align}\label{equ:w04-IN-all}
		2^{-1}\omega_{0, 4} =&\, \left( e^{\hat{t}_{a_1, 0} \angles[\kappa_0, {0, 4}]} \hat{t}_{a_1, 1} \angles[\kappa_1, {0, 4}] + 3 e^{2\hat{t}_{a_1, 0} \angles[\kappa_0, {0, 3}]} \hat{B}_{a_1, 0, a_1, 0}\; \angles[1, {0, 3}]\angles[1, {0, 3}]\right) \prod\nolimits_{j=1}^{4}\dd{\xi_{a_1, 0}(z_j)}\nonumber\\
		& \quad + e^{\hat{t}_{a_1, 0} \angles[\kappa_0, {0, 4}]} \; \angles[\psi, {0, 4}]\; \sum\nolimits_{i=1}^4\left(\dd{\xi_{a_1, 1}(z_i)} \prod\nolimits_{j\neq i}^4 \dd{\xi_{a_1, 0}(z_j)}\right) \nonumber\\
		& + \left( e^{\hat{t}_{a_2, 0} \angles[\kappa_0, {0, 4}]} \hat{t}_{a_2, 1} \angles[\kappa_1, {0, 4}] + 3 e^{2\hat{t}_{a_2, 0} \angles[\kappa_0, {0, 3}]} \hat{B}_{a_2, 0, a_2, 0}\; \angles[1, {0, 3}]\angles[1, {0, 3}]\right) \prod\nolimits_{j=1}^{4}\dd{\xi_{a_2, 0}(z_j)}\nonumber\\
		& \quad + e^{\hat{t}_{a_2, 0} \angles[\kappa_0, {0, 4}]} \; \angles[\psi, {0, 4}]\; \sum\nolimits_{i=1}^4\left(\dd{\xi_{a_2, 1}(z_i)} \prod\nolimits_{j\neq i}^4 \dd{\xi_{a_2, 0}(z_j)}\right)\nonumber\\
		& + e^{(\hat{t}_{a_1,0}+\hat{t}_{a_2,0})\angles[\kappa_0, {0, 3}]}  \; \hat{B}_{a_1, 0, a_2, 0}\;  \angles[1, {0, 3}]\angles[1, {0, 3}]\; \nonumber\\
		& \qquad\qquad\qquad\quad \times \frac{1}{2} \sum\nolimits_{i=1}^4\sum\nolimits_{j\neq i}^4 \dd{\xi_{a_1, 0}(z_i)} \dd{\xi_{a_1, 0}(z_j)} \prod\nolimits_{k\neq i, j}^4 \dd{\xi_{a_2, 0}(z_k)} \nonumber\\
		=&\; \sum_{\sigma\in\{1, 2\}} \left[ e^{2\hat{t}_{a_\sigma, 0} } \left( \hat{t}_{a_\sigma, 1}  + 3  \hat{B}_{a_\sigma, 0, a_\sigma, 0}\; \right) \prod\nolimits_{j=1}^{4}\dd{\xi_{a_\sigma, 0}(z_j)} \right.\nonumber\\
		& \qquad\qquad\qquad \left. + e^{2 \hat{t}_{a_\sigma, 0} } \;  \sum\nolimits_{i=1}^4\left(\dd{\xi_{a_\sigma, 1}(z_i)} \prod\nolimits_{j\neq i}^4 \dd{\xi_{a_1, 0}(z_j)}\right)\; \right] \nonumber\\
		& + e^{(\hat{t}_{a_1,0}+\hat{t}_{a_2,0})}  \; \hat{B}_{a_1, 0, a_2, 0}  \frac{1}{2} \sum\nolimits_{i=1}^4\sum\nolimits_{j\neq i}^4 \dd{\xi_{a_1, 0}(z_i)} \dd{\xi_{a_1, 0}(z_j)} \prod\nolimits_{k\neq i, j}^4 \dd{\xi_{a_2, 0}(z_k)} \, .
	\end{align}
\end{example}

\begin{example}\label{eg:w11-1}
	In the same setting as above, consider $\omega_{1, 1}$. In Example~\ref{ex:mbarc112} it is laid down that $\Mbarc[1, 1, 2] \simeq \Mbar[1, 1] \sqcup \Mbar[1, 1]$. Thus from Theorem~\ref{thm:main-thm} one finds
	\begin{align}
		2^{-1} \omega_{1, 1}(z_1) = \bigintsss_{\Mbar[1, 1]} \hat{\Lambda}_{a_1} \; \hat{B}_{a_1}(z_1; 1/\psi_1) + \bigintsss_{\Mbar[1, 1]} \hat{\Lambda}_{a_2} \; \hat{B}_{a_2}(z_2; 1/\psi_2)\, .
	\end{align}
	Expanding the integrand and keeping only contributions up to degree one, yields for both summands in the above $(\sigma=1, 2)$
	\begin{align}
	 	&\bigintsss_{\Mbar[1, 1]} \left(e^{\hat{t}_{a_\sigma, 0} \kappa_0} (1+\hat{t}_{i, 1}\kappa_1)\right) \left(1+\frac{1}{2}\sum\nolimits_{\delta \in \partial\Mbar[0, 4]} \hat{B}_{a_\sigma, 0, a_\sigma, 0}\psi^0\psi'^0\right) (\dd{\xi_{a_\sigma, 0}(z_1)} \psi_1^0 + \dd{\xi_{a_\sigma, 1}(z_1)} \psi_1^1)\nonumber\\
	 	&= \left( e^{\hat{t}_{a_\sigma, 0} \angles[\kappa_0, {1, 1}]} \hat{t}_{a_\sigma, 1} \angles[\kappa_1, {1, 1}] + \frac{1}{2}e^{\hat{t}_{a_\sigma, 0} \angles[\kappa_0, {0, 3}]} \hat{B}_{a_\sigma, 0, a_\sigma, 0}\; \angles[1, {0, 3}]\right) \dd{\xi_{a_\sigma, 0}(z_1)}\nonumber\\
	 	& \qquad\qquad\qquad + e^{\hat{t}_{a_\sigma, 0} \angles[\kappa_0, {1, 1}]} \; \angles[\psi_1, {1, 1}]\; \dd{\xi_{a_\sigma, 1}(z_1)}\nonumber\\
	 	&= e^{\hat{t}_{a_\sigma, 0}} \left( \frac{\hat{t}_{a_\sigma, 1}}{24}  + \frac{\hat{B}_{a_\sigma, 0, a_\sigma, 0}}{2}\right) \dd{\xi_{a_\sigma, 0}(z_1)}  + e^{\hat{t}_{a_\sigma, 0} } \;\frac{1}{24}\; \dd{\xi_{a_\sigma, 1}(z_1)}\, . \label{equ:w11-IN}
	\end{align}
\end{example}
\newpage
\subsection{Parameters $x_{a, n}$ and $y_{a, n}$}\label{sec:XiYi}
In order to compare the expressions obtained in the previous section to results obtained from topological recursion for physical theories it is convenient to transform to a different equivalent set of moduli of the spectral curve. In the section above, the A-side moduli $t_{a, k}$ and $B_{a, k, a', k'}$ are used to expand the correlators. In the following, we transform to 
\begin{align}
	x_{a, n}= \left. \frac{{\partial_z}^{n+2} x(z)}{{\partial_z}^{2}x(z)} \right|_{z=a}\, , && y_{a, n}= \left. \frac{{\partial_z}^{n+1} y(z)}{\partial_z y(z)} \right|_{z=a}\, ,
\end{align}
for $n\in\NN^\times$, supplemented with $x_{a, 0}=\left.{\partial_z}^{2}x(z)\right|_{z=a}$ and $y_{a, 0}=\left.\partial_z y(z)\right|_{z=a}$. These are well defined, as the spectral curve is required to be regular. However, the definitions can be generalized to a larger class of spectral curves. 
\paragraph{Times $t_{a, k}$ } In order to obtain the times $t_{a, k}$ in terms of the parameters $x_{a, n}$ and $y_{a, n}$, one compares the expansion of $y$ containing $x_{a, n}$ and $t_{a, k}$ to that containing $y_{a, n}$ as
\begin{align}
	y(\{x_{a, n}, t_{a, k}\}; z) = y(\{y_{a, n}\}; z)\, .
\end{align}
The left-hand-side is obtained by expanding first, as explained in equation~\eqref{equ:times}, in terms of $\zeta_a(z)$, and then in terms of $z$ at the ramification point
\begin{align}\label{equ:y-(xit)}
	y(\{x_{a, n}, t_{a, k}\}; z) \overset{z\to a}{=} \sum_{k\geq 0} t_{a, k+2} \, \zeta_a(z)^{k} \overset{z\to a}{=} \sum_{k\geq0} t_{a, k+2} \sum_{l\geq k} \zz_{a, k, l}(\{x_{a, n}\}) (z-a)^l \, . 
\end{align}
There $\zz_{a, k, l}(\{x_{a, n}\})$ is the coefficient of $(z-a)^l$ in the expansion of $(\zeta_{a})^k=(x(z)-x(a))^{k/2}$, which depends on the parameters $\{x_{a, n}\}$. \out[{You can give a general equation for the $\zz_{a, k, l}$, see Appendix~\ref{app:zz-coeffs}.}]\\
The right-hand-side is simply
\begin{align}\label{equ:y-(yi)}
	y(\{y_{a, n}\}; z) \overset{z\to a}{=} y(a) + y_{a, 0}(z-a) +  y_{a, 0} \sum_{n\geq1} y_{a, n}\frac{(z-a)^{n+1}}{(n+1)!}\, .
\end{align}
A comparison of coefficients in equations~\eqref{equ:y-(xit)} and \eqref{equ:y-(yi)} yields relations between the parameters $t_{a, k}$ and $x_{a, n}$ as well as $y_{a, n}$. The expressions for the first few $t_{a, k}$ can be found in Appendix~\ref{app:complexmodel-expansiondata-xiyi}.
\paragraph{Coefficients $B_{a, k; a', k'}$ } A similar analysis that gave the times $t_{a, k}$ in terms of the new set of moduli of the spectral curve, also gives the expansion coefficients $B_{a, k; a', k'}$ in terms of $x_{a, n}$ and $y_{a, n}$. In particular, one obtains the set $\{B_{a, k; a', k'}\}_{k, k'\in\NN}$ by expanding the Bergman kernel in the local coordinate $\zeta_{a}=\sqrt{x(z)-x(a)}$. Again, expanding the appearing powers of $\zeta_{a}(z)$ in $z$ gives
\allowdisplaybreaks
\begin{align}
	B(z, z') &\overset{\substack{z\to a\\z'\to a'}}{=} \left[\frac{\delta_{a, a'}}{(\zeta_a(z)-\zeta_{a'}(z'))^2} + \sum_{k, k'\geq 0} B_{a, k; a', k'}\, \zeta_a(z)^k \zeta_{a'}(z')^{k'}\right] \dd{\zeta_a(z)} \dd{\zeta_{a'}(z')} \\
	&\overset{\substack{z\to a\\z'\to a'}}{=} \dd{z}\dd{z'} \left(\sum_{n\geq0}\sum_{n=0}^m (n+1)(n-m+1)\zz_{a, 1, n+1}\zz_{a', 1, n-m+1}(z-a)^n(z'-a')^{n-m}\right)\nonumber\\
	&\qquad\times \left[ \frac{\delta_{a, a'}}{\left[\sum_{l\geq1} \zz_{a, 1, l}\left((z-a)^l-(z'-a)^l\right)\right]^2}\right.\nonumber\\
	&\qquad\qquad\qquad\qquad \left.+ \sum_{k, k'\geq 0}B_{a, k; a', k'}\sum_{l\geq0}\sum_{r=0}^l \zz_{a, k, r}\zz_{a', k', l-r}(z-a)^r(z'-a')^{l-r}\right]\, , 
\end{align}
with $\zz_{a, k, l}=\zz_{a, k, l}(\{x_{a, n}\})$ as above. This expression is compared with the explicit direct expansion of $B(z, z')$ for $z\to a, z'\to a'$, yielding relations between $\{B_{a, k; a', k'}\}$ and $\{x_{A, n}\}_{A\in\{a, a'\}}$. In Appendix~\ref{app:complexmodel-expansiondata-xiyi} these expressions can be found for the lowest order coefficients.\\

Note that in Theorem~\ref{thm:main-thm} the B-side moduli are used to expand the correlators in terms of intersection numbers on the moduli space of complex curves. However, there are explicit relations between A- and B-side moduli given in equations~\eqref{equ:t-ttilde} and \eqref{equ:B-hatB}.\\\\
In the following, the expressions obtained in Examples~\ref{eg:w04-1} and \ref{eg:w11-1} are rewritten in terms of the new set of moduli of the spectral curve.
\begin{example}\label{eg:w04-2}
	In the discussion of $\omega_{0, 4}$ in Example~\ref{eg:w04-1} its expansion in terms of intersection numbers on the moduli space of complex curves was re-derived. Using the relation of $t_{a, k}$ and $B_{a, k; a', k'}$ to $x_{a, n}$ and $y_{a, n}$ established above and explicitly written for the first few parameters in Appendix~\ref{app:complexmodel-expansiondata-xiyi}, the expansion given in equation~\eqref{equ:w04-IN-all} is rewritten to obtain
	\begin{align}
		2^{-1} \omega_{0, 4} =& \sum_{\sigma\in\{1, 2\}} \frac{1}{48x_{a_\sigma, 0}^2y_{a_\sigma, 0}^2} \left[\frac{(3 - 5 \angles[\kappa_1, {0, 4}] + 20 \angles[\psi, {0, 4}]) x_{a_\sigma, 1}^2 - 3 (1 - \angles[\kappa_1, {0, 4}] + 4 \angles[\psi, {0, 4}]) x_{a_\sigma, 2} }{(z_1-a_\sigma)^2(z_2-a_\sigma)^2(z_3-a_\sigma)^2(z_4-a_\sigma)^2}\right.\nonumber\\
		&\left.\qquad\qquad\qquad\qquad\qquad\qquad\qquad\quad +\frac{12 \angles[\kappa_1, {0, 4}] x_{a_\sigma, 1} y_{a_\sigma, 1} - 12 \angles[\kappa_1, {0, 4}] y_{a_\sigma, 2}}{(z_1-a_\sigma)^2(z_2-a_\sigma)^2(z_3-a_\sigma)^2(z_4-a_\sigma)^2} \right.\nonumber\\
		&\left. \qquad\qquad\qquad\qquad\qquad- 24 \angles[\psi, {0, 4}] x_{a_\sigma, 1} \sum\nolimits_{i=1}^4\left(\frac{1}{(z_i-a_\sigma)^3} \prod\nolimits_{j\neq i}^4 \frac{1}{(z_j-a_\sigma)^2}\right)\right.\nonumber\\
		&\left. \qquad\qquad\qquad\qquad\qquad+ 72 \angles[\psi, {0, 4}] \sum\nolimits_{i=1}^4\left(\frac{1}{(z_i-a_\sigma)^4} \prod\nolimits_{j\neq i}^4 \frac{1}{(z_j-a_\sigma)^2}\right)\right]\nonumber\\
		&+\frac{1/2}{(a_1 - a_2)^2 x_{a_1, 0} x_{a_2, 0} y_{a_1, 0} y_{a_2, 0}}\left( \frac{1}{(z_1-a_1)^2 (z_2-a_1)^2 (z_3-a_2)^2 (z_4-a_2)^2}\right.\nonumber\\
		&\qquad\qquad\qquad\qquad\qquad\qquad\qquad\qquad\qquad\;\left.\white \frac{1}{(a_1)^2}\black + \text{symm}(z_1, \dots, z_4)\right)\, .
	\end{align}
\end{example}
\begin{example}
	In the same way as in the previous example, here, the expansion of $\omega_{1, 1}$ in terms of intersection numbers on the moduli space of complex curves (see equation~\eqref{equ:w11-IN}) is rewritten using $\{x_{a, n}\}$ and $\{y_{a, n}\}$. Using the data presented in Appendix~\ref{app:complexmodel-expansiondata-xiyi} one finds
	\begin{align}
		2^{-1}\omega_{1, 1} 
		=&\; \sum_{\sigma\in\{1, 2\}} \frac{1}{x_{a_\sigma, 0}y_{a_\sigma, 0}}\left[\frac{(-1 + 10 (\angles[\kappa_1, {1, 1}] - \angles[\psi, {1, 1}])) x_{a, 1}^2 + (1 - 6 (\angles[\kappa_1, {1, 1}] - \angles[\psi, {1, 1}])) x_{a, 2}}{48(z_1- a_\sigma)^2}\right.\nonumber\\
		&\; \qquad\qquad\qquad\qquad\qquad\qquad\qquad\qquad-\frac{ 24 \angles[\kappa_1, {1, 1}] x_{a, 1} y_{a, 1} - 24 \angles[\kappa_1, {1, 1}] y_{a, 2}}{48(z_1- a_\sigma)^2}\nonumber\\
		& \qquad\qquad\qquad\qquad\left.+ \frac{\angles[\psi, {1, 1}] x_{a, 1}}{(z_1-a_\sigma)^3} -\frac{3 \angles[\psi, {1, 1}]}{(z_1-a_\sigma)^4}\right]\, .
	\end{align}
\end{example}

\section{The LSZ-model}\label{sec:LSZ}
An approach to the study of quantum field theories on non-commutative space is provided by the expansion of the quantum field in terms of a matrix base obeying the Moyal algebra. Then properties can be studied in the finite dimensional approximation and limiting procedures to recover the quantum field theory limit have to be carried out carefully. The partition function of the LSZ model in the finite dimensional approximation of complex matrices of size $N$ is given by 
\begin{align}\label{equ:partitionfunc-LSZ}
	\Zz= \int_{M_N} \dd{\Phi}\dd{\Phi^\dagger}\; \exp(-N\tr[E\, \Phi\Phi^\dagger+\tilde{E}\, \Phi^\dagger\Phi+\frac{\lambda}{2}\Phi^\dagger\Phi\Phi^\dagger\Phi])\, . 
\end{align}
In the above, $E$ and $\tilde{E}$ are distinct hermitian matrices of size $N$ with positive distinct eigenvalues $\{E_1, \dots, E_N\}$ and $\{\tilde{E}_1, \dots, \tilde{E}_N\}$. Note that these can be assumed to be diagonal due to the invariance of the partition function under unitary transformations. The coupling strength is given by $\lambda\in\CC$. The above partition function can also be considered a complex analogue of the quartic Kontsevich model. In contrast to the present LSZ model, the more constrained setting of the hermitian model results in a more complicated solution structure.\\
In the original work \cite{Langmann_2004} the authors explicitly computed some correlation functions in the case of $\tilde{E}=0$ and proved integrability in the sense of Toda. Recently, in \cite{BH2205} this result has been generalized to all external fields and topological recursion provides the machinery to compute arbitrary high correlation functions. Furthermore, it was specified that the partition function~\eqref{equ:partitionfunc-LSZ} is in fact a $\tau$-function of the Kadomtsev-Petviashvili hierarchy (for $N<\infty$), a subclass of the Toda hierarchy. It was shown that the model is entirely described by the spectral curve $\Ss_{\text{LSZ}}=(\PP^1, x_{\tilde{d}}, y_d, B)$ with 
\begin{align}
	x_{\tilde{d}}(z) = z- \frac{\lambda}{N} \sum_{k=1}^{\tilde{d}} \frac{\tilde{r}_k}{y'(\tilde{\eps}_k)(z-\tilde{\eps}_k)}\, ,&&   y_d(z) = -z+ \frac{\lambda}{N} \sum_{k=1}^{d} \frac{r_k}{x'(\eps_k)(z-\eps_k)}\, ,
\end{align}
\begin{align}
	B(z_1, z_2)=\frac{\dd{z_1}\dd{z_2}}{(z_1-z_2)^2}\, .
\end{align}
In the above the points $\eps_k$ and $\tilde{\eps}_k$ are defined by $e_k=x(\eps_k)$ and $\tilde{e}_k=x(\tilde{\eps}_k)$, respectively, where $\{e_k\}_{k=1, \dots, d}$ and $\{\tilde{e}_k\}_{k=1, \dots, \tilde{d}}$ are pairwise distinct eigenvalues of $E$ and $\tilde{E}$ with multiplicities $r_k$ and $\tilde{r}_k$ summing up to $N$.\\
In this work we restrict the analysis to the case of $d=\tilde{d}=1$. Then, the above functions $x$ and $y$ can be reduced to 
\begin{align}
	x(z)= z + \frac{\gamma^2}{y'(\tilde{\eps}) (z - \tilde{\eps})}\, , && y(z)= -z + \frac{\gamma^2}{x'(\eps)(-z + \eps)}\, ,
\end{align}
with $\eps=\eps_1$ and $\tilde{\eps}=\tilde{\eps}_1$ as well as $\gamma^2=-\lambda r_1/N=-\lambda\tilde{r}_1/N=-\lambda$. Note that also the implicit definitions of $\eps$ and $\tilde{\eps}$ in terms of $e$ and $e$ as well as the derivative of $y$ in the definition of $x$ and vice versa can be made explicit. \\
This setting is often referred to as the combinatorial limit, as it has been subject of studies already in the 1990s in the field of combinatorics and enumeration. Notable works include \cite{MORRIS1991703, ZJZ03}. In particular, this model (with arbitrary polynomial interaction) is known to count bipartite maps\footnote{For a nice introduction to maps and graphs the reader might be referred to \cite{Eynard:2016yaa}.}. These are ordinary maps of even face degree with black or white colored vertices and no monochromatic edge. An edge as well as its marking plus the orientation of an edge creates a rooted bipartite graph and a boundary of even length $2l_k$ following the face to the right of the rooted edge. As long as they do not correspond to the same boundary, also multiple edges can be marked. Denoting by $\Tt^{(g)}_{2l_1, \dots, 2l_n}$ the generating function of bipartite maps with natural embedding into Riemann surfaces of genus $g$ and $n$ boundaries of length $2l_1, \dots, 2l_n$, one can relate the counting results to the correlators of topological recursion through
\begin{align}
	&\Tt^{(g)}_{2l_1, \dots, 2l_n} = (-1)^n \Res[z_1, \dots, z_n\to\infty] x(z_1)^{l_1}\cdots x(z_n)^{l_n} \omega_{g, n}(\Ss_\text{LSZ}; z_1, \dots, z_n)\nonumber\\
	&\Leftrightarrow \qquad \omega_{g, n}(\Ss_\text{LSZ}; z_1, \dots, z_n) = \sum_{l_1, \dots, l_n =1}^{\infty} \frac{\Tt^{(g)}_{2l_1, \dots, 2l_n}}{x(z_1)^{l_1+1}\cdots x(z_n)^{l_n+1}}\dd{x(z_1)}\cdots\dd{x(z_n)}\, .
\end{align}
Remark that comparing this model (with quartic interaction) to its hermitian counterpart, that is the quartic Kontsevich model, the latter counts \textit{ordinary} rooted quadrangulations while the quartic LSZ model counts \textit{bipartite} ones. This, again, stresses the relation of both models, which is to be exploited in the further understanding of the quartic Kontsevich model in higher genus.
\newpage
\section{Result and Outlook}\label{sec:results}
The aim of this work is to verify the agreement of correlation functions calculated directly from topological recursion (Section~\ref{sec:topological-recursion}) and expressed in terms of intersection numbers on the moduli space of curves (Section~\ref{sec:TR-IN}), which is provided by \cite{Eynard1110} as a theorem, in the concrete case of the LSZ model in the combinatorial limit. \\
Therefore, the correlation functions $\omega_{0, 4}$ and $\omega_{1, 1}$ are considered here. These are recursively calculated by computer algebra. Furthermore, the expansion data listed in Appendix~\ref{app:complexmodel-expansiondata-xiyi} is used to specify Section~\ref{sec:TR-IN} to the LSZ model in the combinatorial limit with quartic interaction, as explained in Section~\ref{sec:XiYi}. Subtracting one expression from the other and evaluating the (higher) residues at the poles, which are only at the ramification points of the spectral curve, verifies that both expressions give the same, i.e. for all $a_i \in \{\text{ram.pts.}\}$, $i\in \{1, 2, \dots , n\}$, and all $\alpha\in \NN^n$
\begin{align}\label{equ:TR-IN}
		[(\vec{z}-\vec{a})^{-\alpha}]\left(\omega_{g, n}^\text{TR}(\vec{z})-\omega_{g, n}^\text{IN-TR}(\vec{z})\right) = 0 \, ,
\end{align}
for $(g, n)\in\{(0, 4), (1, 1)\}$. Note that, beyond $g=1$, the expressions $\omega^\text{IN-TR}$ get lengthy and clutter the analysis -- beyond $n=4$ in genus zero, the pole structure gets more and more involved. \\
By the study of these first examples, the general technique of comparing the two expressions in equation~\eqref{equ:TR-IN} is shown, and different conventions in the literature are clarified. This will enable further studies of the intersection theory encoded in physical models such as in the quartic Kontsevich model. This model is conjectured to be governed by a consistent extension of topological recursion, called blobbed topological recursion \cite{Borot:2015hna}. The correlation functions generated by this extension have contributions that are holomorphic at the ramification points of the spectral curve. Nevertheless, there is a theory (provided in Section~3 of \cite{Borot:2015hna}) of how to expand these in terms of intersection numbers on the moduli space of curves. Furthermore, there is evidence that the genus zero sector of the quartic Kontsevich model can be extracted from the LSZ-model. To utilize the explicit expressions found in here for the analysis of (the planar sector of the) quartic Kontsevich model will be subject of future work. 

\section*{Acknowledgments}
This work is funded by the Deutsche Forschungsgemeinschaft through the Research Training Group 2149 "GRK 2149: Strong and Weak Interactions -- from Hadrons to Dark Matter"  as well as supported\footnote{funded by the Deutsche Forschungsgemeinschaft (DFG, German Research Foundation) under Germany's Excellence Strategy EXC 2044 –390685587, Mathematics Münster: Dynamics–Geometry–Structure} by the Cluster of Excellence Mathematics Münster.

\newpage
\begin{appendices}

\section{Expansion data of the combinatorial limit of the LSZ model with quartic interaction} \label{app:complexmodel-expansiondata-xiyi}
In the following, the expansion data of the spectral curve associated to the complex quartic Kontsevich model are provided. Remark that for some models it is possible to give the expansion coefficients $\hat{t}_{a, k}$ and $\hat{B}_{a, k; a', k'}$ for all $k$ and $k'$ in closed form\footnote{Instructive examples can for example be found in \cite{Eynard1104}.}. This is not possible here, as the integrals in equations~\eqref{equ:t-tildet} and \eqref{equ:hatB} for this model do not evaluate into some known functions. Therefore, the coefficients $\hat{t}_{a, k}$ and $\hat{B}_{a, k; a', k'}$ are expressed in terms of $\{x_{a, n}\}$ and $\{y_{a, n}\}$ defined in Section~\ref{sec:XiYi}. A list of the first coefficients is provided in Tables~\ref{tab:t-xiyi}--\ref{tab:Ba1a2-xiyi}. \\
Recall, the spectral curve contains the information about the two functions
\begin{align}
	x(z)= z + \frac{\gamma_y^2}{z - \tilde{\eps}} && y(z)= -z + \frac{\gamma_x^2}{-z + \eps}\, ,
\end{align}
with $\gamma_y^2 =-\lambda/y'(\tilde{\eps})$ and $\gamma_x^2= -\lambda/x'(\eps)$. The ramification points of this spectral curve defined by $\dd{x(a)}=0$ are at
\begin{align}
	a_\pm= \tilde{\eps} \pm \gamma_y\, .
\end{align}
Using this information the moduli of the spectral curve $x_{a, n}$ and $y_{a, n}$ can be specified. One finds
\begin{align}
	x_{a_\pm, n>0}  = \frac{(n+2)!}{2(\mp\gamma_y)^{n}}\, , && y_{a_\pm, n>0}  = \frac{(n+1)!}{\left(\eps - \tilde{\eps}\mp \gamma_y\right)^{n} - \gamma_x^{-2} \left( \eps-\tilde{\eps} \mp \gamma_y \right)^{n+2}}\, ,
\end{align}
as well as
\begin{align}
	x_{a_\pm, 0} = \pm 2\gamma_y^{-1} \, , && y_{a_\pm, 0} = -1+ \gamma_x^{2}(\eps-\tilde{\eps}\mp \gamma_y)^{-2}\, .
\end{align}
{\begin{table}[H]
	\centering
	\caption{This lists the first expansion coefficients or \textit{times} $t_{a, k}$ of $y(z)$ at the ramification point $z=a$ expressed in terms of the parameters $x_{a, n}$ and $y_{a, n}$.}
	\label{tab:t-xiyi}
	\begin{tabular}{c||c}
		$k$ & $t_{a, k}$ \\
		\hline\hline
		3 & $\substack{\frac{y_{a, 0}}{x_{a, 0}^{1/2}}\sqrt{2}}$ \\
		4 & $\substack{\frac{y_{a, 0}}{3x_{a, 0}}\left(-x_{a,1}+3y_{a,1}\right)}$ \\
		5 & $\substack{\frac{y_{a, 0}}{18 \sqrt{2}} \left(-12 x_{a,1} y_{a,1}+5 x_{a,1}^2-3 x_{a,2}+12 y_{a,2}\right)}$ \\
		6 & $\substack{\frac{y_{a, 0}}{270x_{a, 0}^{3/2}} \left(90 x_{a,1}^2 y_{a,1}-90 x_{a,1} y_{a,2}-45 x_{a,2} y_{a,1}-40 x_{a,1}^3+45 x_{a,2} x_{a,1}-9 x_{a,3}+45 y_{a,3}\right)}$ \\
		7 & $\substack{\frac{y_{a, 0}}{2160 \sqrt{2}x_{a, 0}^2}\left(-840 x_{a,1}^3 y_{a,1}+840 x_{a,1}^2 y_{a,2}+840 x_{a,2} x_{a,1} y_{a,1}-480 x_{a,1} y_{a,3}-144 x_{a,3} y_{a,1}-360 x_{a,2} y_{a,2}+385 x_{a,1}^4\right.\\
			\qquad\qquad\qquad\qquad\qquad\qquad\qquad\left.-630 x_{a,2} x_{a,1}^2+168 x_{a,3} x_{a,1}+105 x_{a,2}^2-24 x_{a,4}+144 y_{a,4}\right)}$ \\
		8 & $\substack{\frac{y_{a, 0}}{17010x_{a, 0}^{5/2}} \left(4200 x_{a,1}^4 y_{a,1}-4200 x_{a,1}^3 y_{a,2}-6300 x_{a,2} x_{a,1}^2 y_{a,1}+2520 x_{a,1}^2 y_{a,3}+1512 x_{a,3} x_{a,1} y_{a,1}+3780 x_{a,2} x_{a,1} y_{a,2}\right.\\
			\left.-945 x_{a,1} y_{a,4}+945 x_{a,2}^2 y_{a,1}-189 x_{a,4} y_{a,1}-567 x_{a,3} y_{a,2}-945 x_{a,2} y_{a,3}-1960 x_{a,1}^5+4200 x_{a,2} x_{a,1}^3\right.\\
			\qquad\qquad\qquad\qquad\left.-1260 x_{a,3} x_{a,1}^2-1575 x_{a,2}^2 x_{a,1}+252 x_{a,4} x_{a,1}+378
		x_{a,2} x_{a,3}-27 x_{a,5}+189 y_{a,5} \right)}$ \\
		9 & $\substack{\frac{y_{a, 0}}{2721600 \sqrt{2}x_{a, 0}^3}\left(-900900 x_{a,1}^5 y_{a,1}+900900 x_{a,1}^4 y_{a,2}+1801800 x_{a,2} x_{a,1}^3 y_{a,1}-554400 x_{a,1}^3 y_{a,3} -498960 x_{a,3} x_{a,1}^2 y_{a,1}\right.\\
			\left.\qquad\qquad\quad-1247400 x_{a,2} x_{a,1}^2 y_{a,2}+226800 x_{a,1}^2 y_{a,4}-623700 x_{a,2}^2 x_{a,1} y_{a,1}+90720 x_{a,4} x_{a,1} y_{a,1}+272160 x_{a,3} x_{a,1} y_{a,2}\right.\\
			\left.\qquad+453600 x_{a,2} x_{a,1} y_{a,3}-60480 x_{a,1} y_{a,5}+136080 x_{a,2} x_{a,3}y_{a,1}-8640 x_{a,5} y_{a,1}+170100 x_{a,2}^2 y_{a,2}\right.\\
			\left.\qquad\qquad-30240 x_{a,4} y_{a,2}-60480 x_{a,3} y_{a,3}-75600 x_{a,2} y_{a,4}+425425 x_{a,1}^6-1126125 x_{a,2} x_{a,1}^4+360360 x_{a,3} x_{a,1}^3\right.\\
			\left.+675675 x_{a,2}^2 x_{a,1}^2-83160 x_{a,4} x_{a,1}^2-249480 x_{a,2}x_{a,3} x_{a,1}+12960 x_{a,5} x_{a,1}-51975 x_{a,2}^3\right.\\
			\qquad\qquad\qquad\qquad\qquad\qquad\qquad\qquad\left.+13608 x_{a,3}^2+22680 x_{a,2} x_{a,4}-1080 x_{a,6}+8640 y_{a,6}\right)}$
	\end{tabular}
\end{table}
\begin{table}[H]
	\centering
	\caption{This lists the first expansion coefficients or \textit{dual times} $\hat{t}_{a, k}$ of the Laplace transform of $\dd{y(z)}$ at the ramification point $z=a$ in terms of the parameters $x_{a, n}$ and $y_{a, n}$. These coefficients are dual to the  times $t_{a, k}$ (see Table~\eqref{tab:t-xiyi}) and their relation is provided in the second column.}
	\begin{tabular}{c||c|c}
	
	$k$ & \multicolumn{2}{c}{$\hat{t}_{a, k}$} \\
	\hline \hline
	$0$ & $\log \frac{1}{2 t_{a,3}}$ & $\substack{\log \left(\frac{\sqrt{x_{a,0}}}{2 \sqrt{2} y_{a,0}}\right)}$ \\
	\hline
	$1$ & $-\frac{3 t_{a,5}}{2 t_{a,3}}$ & $\substack{\frac{1}{24x_{a, 0}}\left(12 x_{a,1} y_{a,1}-5 x_{a,1}^2+3 x_{a,2}-12 y_{a,2}\right)}$\\
	\hline
	$2$ & $\substack{\frac{9 t_{a,5}^2}{8 t_{a,3}^2}\\-\frac{15 t_{a,7}}{4 t_{a,3}}}$ & $\substack{\frac{1}{48x_{a, 0}^2}\left(30 x_{a,1}^3 y_{a,1}+6 x_{a,1}^2 y_{a,1}^2-30 x_{a,1}^2 y_{a,2}-32 x_{a,2} x_{a,1} y_{a,1}-12 x_{a,1} y_{a,1} y_{a,2}+20 x_{a,1}y_{a,3}\right.\\
		\qquad\qquad\left.+6 x_{a,3} y_{a,1}+12 x_{a,2} y_{a,2}-15 x_{a,1}^4+25 x_{a,2} x_{a,1}^2-7 x_{a,3} x_{a,1}-4 x_{a,2}^2+x_{a,4}+6 y_{a,2}^2-6 y_{a,4}\right)}$\\
	\hline
	$3$ & $\substack{-\frac{9 t_{a,5}^3}{8 t_{a,3}^3}\\+\frac{45 t_{a,7} t_{a,5}}{8t_{a,3}^2}\\-\frac{105 t_{a,9}}{8 t_{a,3}} }$ & $\substack{\frac{1}{5760x_{a, 0}^3} \left(10800 x_{a,1}^5 y_{a,1}+1800 x_{a,1}^4 y_{a,1}^2-10800 x_{a,1}^4 y_{a,2}+240 x_{a,1}^3 y_{a,1}^3-22200 x_{a,2} x_{a,1}^3 y_{a,1}\right.\\
		\qquad\qquad\left.-3600x_{a,1}^3 y_{a,1} y_{a,2}+7200 x_{a,1}^3 y_{a,3}-1920 x_{a,2} x_{a,1}^2 y_{a,1}^2+1800 x_{a,1}^2 y_{a,2}^2\right.\\
		\qquad\qquad\left.+6360 x_{a,3} x_{a,1}^2y_{a,1}-720 x_{a,1}^2 y_{a,1}^2 y_{a,2}+15000 x_{a,2} x_{a,1}^2 y_{a,2}+1200 x_{a,1}^2 y_{a,1} y_{a,3}\right.\\
		\qquad\qquad\left.-3000 x_{a,1}^2 y_{a,4}+360 x_{a,3}x_{a,1} y_{a,1}^2+720 x_{a,1} y_{a,1} y_{a,2}^2+7920 x_{a,2}^2 x_{a,1} y_{a,1}\right.\\
		\qquad\qquad\left.-1200 x_{a,4} x_{a,1} y_{a,1}-3360 x_{a,3} x_{a,1}y_{a,2}+2640 x_{a,2} x_{a,1} y_{a,1} y_{a,2}-6000 x_{a,2} x_{a,1} y_{a,3}\right.\\
		\qquad\qquad\left.-1200 x_{a,1} y_{a,2} y_{a,3}-360 x_{a,1} y_{a,1} y_{a,4}+840x_{a,1} y_{a,5}-720 x_{a,2} y_{a,2}^2-1800 x_{a,2} x_{a,3} y_{a,1}\right.\\
		\qquad\qquad\left.+120 x_{a,5} y_{a,1}-1920 x_{a,2}^2 y_{a,2}+360 x_{a,4} y_{a,2}-360x_{a,3} y_{a,1} y_{a,2}+840 x_{a,3} y_{a,3}\right.\\
		\qquad\qquad\left.+960 x_{a,2} y_{a,4}-5525 x_{a,1}^6+14775 x_{a,2} x_{a,1}^4-4830 x_{a,3} x_{a,1}^3-8900x_{a,2}^2 x_{a,1}^2+1130 x_{a,4} x_{a,1}^2\right.\\
		\qquad\qquad\left.+3360 x_{a,2} x_{a,3} x_{a,1}-180 x_{a,5} x_{a,1}+660 x_{a,2}^3-189 x_{a,3}^2-300 x_{a,2}x_{a,4}+15 x_{a,6}-240 y_{a,2}^3\right.\\
		\qquad\qquad\qquad\qquad\qquad\qquad\qquad\qquad\qquad\qquad\qquad\quad\left.+360 y_{a,2} y_{a,4}-120 y_{a,6}\right)}$ \\
	\end{tabular}
	\label{tab:ttilde-xiyi}
\end{table}
\begin{table}[H]
	\centering
	\caption{This lists the expansion coefficients $B_{a, k; a, k'}$ of $B(z, z')$ at the ramification point $(z, z')=(a, a)$ in the local variables $\zeta_{a}(z)=\sqrt{x(z)-x(a)}$. Note that the dual coefficients $\hat{B}_{a, 2k; a, 2k'}$ can be found be equation~\eqref{equ:B-hatB} and differ only by a numerical factor.}
	\label{tab:Baa-xiyi}
	\begin{tabular}{c||>{\centering\arraybackslash}p{3cm}>{\centering\arraybackslash}p{4.3cm}c}
		\diagbox{k}{k'}&0&1&2\\
		\hline\hline
		0 & $\frac{1}{12x_{a, 0}}\left(x_{a,1}^2-x_{a,2}\right)$ & $\substack{\frac{1}{135x_{a, 0}^{3/2}}\left(-10 \sqrt{2} x_{a,1}^3\right. \\ \qquad\left. +15 \sqrt{2} x_{a,2} x_{a,1}-\frac{9 x_{a,3}}{\sqrt{2}}\right)}$ & $\substack{\frac{1}{1440x_{a, 0}^2}\left(175 x_{a,1}^4\right. \\ \qquad\left.-350 x_{a,2} x_{a,1}^2+120 x_{a,3} x_{a,1}\right. \\ \qquad\qquad\left.+75 x_{a,2}^2-24 x_{a,4}\right)\\ \;}$ \\
		1 & $\substack{\frac{1}{135x_{a, 0}^{3/2}}\left(-10 \sqrt{2} x_{a,1}^3\right.\qquad \\ \left. +15 \sqrt{2} x_{a,2} x_{a,1}-\frac{9 x_{a,3}}{\sqrt{2}}\right)}$ & $\substack{\frac{1}{270x_{a, 0}^2}\left(40 x_{a,1}^4\right. \\ \qquad\left. -80 x_{a,2} x_{a,1}^2+30 x_{a,3} x_{a,1}\right. \\ \qquad\qquad\left. +15 x_{a,2}^2-6 x_{a,4}\right)}$ & $\substack{\frac{1}{3780x_{a, 0}^{5/2}}\left(-490 \sqrt{2} x_{a,1}^5\right.\qquad\qquad \\ \qquad\left.+1225 \sqrt{2} x_{a,2} x_{a,1}^3-\frac{945 x_{a,3} x_{a,1}^2}{\sqrt{2}}\right. \\ \qquad\left.-525 \sqrt{2} x_{a,2}^2 x_{a,1}+126 \sqrt{2} x_{a,4} x_{a,1}\right. \\ \qquad\qquad\left.+\frac{315 x_{a,2} x_{a,3}}{\sqrt{2}}-18 \sqrt{2} x_{a,5}\right)\\ \;}$ \\
		2 & $\substack{\frac{1}{1440x_{a, 0}^2}\left(175 x_{a,1}^4\right.\qquad\qquad \\ \left.-350 x_{a,2} x_{a,1}^2+120 x_{a,3} x_{a,1}\right. \\ \left.+75 x_{a,2}^2-24 x_{a,4}\right)}$ & $\substack{\frac{1}{3780x_{a, 0}^{5/2}}\left(-490 \sqrt{2} x_{a,1}^5\right.\qquad\qquad \\ \qquad\left.+1225 \sqrt{2} x_{a,2} x_{a,1}^3-\frac{945 x_{a,3} x_{a,1}^2}{\sqrt{2}}\right. \\ \qquad\left.-525 \sqrt{2} x_{a,2}^2 x_{a,1}+126 \sqrt{2} x_{a,4} x_{a,1}\right. \\ \qquad\qquad\left.+\frac{315 x_{a,2} x_{a,3}}{\sqrt{2}}-18 \sqrt{2} x_{a,5}\right)}$ & $\substack{\frac{1}{181440 x_{a, 0}^{5/2}}\left(42875 x_{a,1}^6\right.\qquad \\ \qquad\left.-128625 x_{a,2} x_{a,1}^4+50400 x_{a,3} x_{a,1}^3\right. \\ \qquad\left.+86625 x_{a,2}^2 x_{a,1}^2-15120	x_{a,4} x_{a,1}^2\right. \\ \qquad\left.-37800 x_{a,2} x_{a,3} x_{a,1}+3024 x_{a,5} x_{a,1}\right. \\ \qquad\left.-7875 x_{a,2}^3+2268 x_{a,3}^2+4536 x_{a,2} x_{a,4}\right. \\ \qquad\qquad\qquad\qquad\left.-324 x_{a,6}\right)}$\\
	\end{tabular}
\end{table}
\begin{table}[H]
	\centering
	\caption{This lists the expansion coefficients $B_{a_1, k; a_2, k'}$ of $B(z, z')$ at the ramification point $(z, z')=(a_1, a_2)$ in the local variables $\zeta_{a_i}(z)=\sqrt{x(z)-x(a_i)}$ omitting a factor $\Delta^{-(k+k'+2)}$. Here $\Delta=a_1-a_2$. Note that the dual coefficients $\hat{B}_{a_1, k; a_2, k'}$ can be found from equation~\eqref{equ:B-hatB} and differ only by a numerical factor.}
	\label{tab:Ba1a2-xiyi}
	\begin{tabular}{c||>{\centering\arraybackslash}p{3.2cm}>{\centering\arraybackslash}p{4.3cm}c}
		\diagbox{k}{k'}&0&1&2\\
		\hline\hline
		0 & $\frac{2}{\sqrt{x_{a_1,0}x_{a_2,0}}}$ & $\substack{-\frac{2\sqrt{2}}{3\sqrt{x_{a_1, 0}}x_{a_2, 0}}\left(\Delta x_{a_2,1}\right.\\ \qquad\qquad\qquad\left.-6\right)}$ & $\substack{\frac{1}{6 \sqrt{x_{a_1, 0}}x_{a_2, 0}^{3/2}} \left(5 \Delta^2 x_{a_2,1}^2\right.\qquad\; \\ \left.-3 \Delta^2 x_{a_2,2}-24 \Delta x_{a_2,1}\right. \\ \qquad\qquad\left.+72\right)}$ \\
		1 & $\substack{-\frac{2\sqrt{2}}{3x_{a_1, 0}\sqrt{x_{a_2, 0}}}\left(\Delta x_{a_1,1}\right.\\ \qquad\qquad\qquad \left. +6 \right)}$ & $\substack{\frac{1}{9 x_{a_1, 0} x_{a_2, 0}}\left(4 \Delta^2 x_{a_1,1} x_{a_2,1}\right.\qquad\\ \left.-24 \Delta x_{a_1,1}+24 \Delta x_{a_2,1}\right.\\ \qquad\qquad\qquad \left.-216\right)}$ & $\substack{\frac{1}{9\sqrt{2} x_{a_1, 0} x_{a_2, 0}^{3/2}} \left(-5 \Delta^3 x_{a_1,1} x_{a_2,1}^2 \right.\qquad \\ \quad\left. +3 \Delta^3 x_{a_1,1} x_{a_2,2}-30 \Delta^2 x_{a_2,1}^2 \right. \\ \quad\left. +24 \Delta^2 x_{a_1,1} x_{a_2,1}  +18 \Delta^2 x_{a_2,2} \right. \\ \quad\left. -72 \Delta x_{a_1,1}  +216 \Delta x_{a_2,1} \right. \\ \qquad\qquad\qquad\left. -864\right)}$ \\
		2 & $\substack{\frac{1}{6 x_{a_1, 0}^{3/2} \sqrt{x_{a_2, 0}}}\left(5 \Delta^2 x_{a_1,1}^2\right.\qquad\; \\ \left.-3 \Delta^2 x_{a_1,2}+24 \Delta x_{a_1,1}\right. \\ \qquad\qquad\left.+72\right)}$ & $\substack{\frac{1}{9\sqrt{2} x_{a_1, 0}^{3/2} x_{a_2, 0}}\left(-5 \Delta^3 x_{a_1,1}^2 x_{a_2,1}\right. \\ \quad\left.+3 \Delta^3 x_{a_1,2} x_{a_2,1}+30 \Delta^2 x_{a_1,1}^2 \right. \\ \quad\left. -18 \Delta^2 x_{a_1,2}-24 \Delta^2 x_{a_1,1} x_{a_2,1} \right. \\ \quad\left. +216 \Delta x_{a_1,1}-72 \Delta x_{a_2,1}\right. \\ \qquad\qquad\qquad\left. +864\right)}$ & $\substack{\frac{1}{72 x_{a_1, 0}^{3/2} x_{a_2, 0}^{3/2}}\left(25 \Delta^4 x_{a_1,1}^2 x_{a_2,1}^2 \right.\qquad\quad \\ \quad\left. -15 \Delta^4 x_{a_1,2} x_{a_2,1}^2-15 \Delta^4 x_{a_1,1}^2 x_{a_2,2} \right. \\ \quad\left. +9 \Delta^4 x_{a_1,2} x_{a_2,2}+120 \Delta^3 x_{a_1,1} x_{a_2,1}^2 \right. \\ \quad\left. -120 \Delta^3 x_{a_1,1}^2x_{a_2,1}+72 \Delta^3 x_{a_1,2} x_{a_2,1} \right. \\ \quad\left. -72 \Delta^3 x_{a_1,1} x_{a_2,2}+360 \Delta^2 x_{a_1,1}^2 \right. \\ \quad\left. +360 \Delta^2 x_{a_2,1}^2-216 \Delta^2 x_{a_1,2} \right. \\ \quad\left. -864 \Delta^2 x_{a_1,1} x_{a_2,1}-216 \Delta^2 x_{a_2,2} \right. \\ \quad\left. +3456 \Delta x_{a_1,1}-3456 \Delta x_{a_2,1} \right. \\ \qquad\qquad\qquad\left. +17280\right)}$ \\
	\end{tabular}
\end{table}}

\end{appendices}

\clearpage

\addcontentsline{toc}{section}{References}

\bibliographystyle{JHEP}
\bibliography{main_bib}

\end{document}